\documentclass[aps,prd,final,twocolumn,letterpaper]{revtex4-1}

\usepackage{graphicx}                      
\usepackage{epstopdf}
\usepackage{amsmath} 
\usepackage{braket}

\newcommand{\rds}{R_{D^*}}
\newcommand{\rdsp}{R_{D\pi}}
\newcommand{\pds}{p_{D^*}}
\newcommand{\dsm}{m_{D^*}}

\begin{document}

\title{Closing the Gap on $R_{D^*}$ by including longitudinal effects}
\author{J. E. Chavez-Saab}
\author{Genaro Toledo}
\affiliation{Instituto de Fisica, Universidad Nacional Autonoma de Mexico, AP20-364, Ciudad de Mexico 01000, Mexico.}
\date{\today} 

\begin{abstract}
\noindent	
Measurements of the $R_{D^*}\equiv\mathrm{Br}(B\rightarrow \tau\nu D^*)/\mathrm{Br}(B\rightarrow e\nu D^*)$ parameter remain in tension with the standard model prediction, despite recent results helping to close the gap. The standard model prediction it is compared with considers the $D^*$ as an external particle, even though what is detected in experiments is a $D\pi$ pair it decays into, from which it is reconstructed. We argue that the experimental result must be compared with the theoretical prediction considering the full 4-body decay $(B\rightarrow l\nu D^* \to l\nu D\pi)$. We show that the longitudinal degree of freedom of the off-shell $D^*$ helps to further close the disagreement gap with experimental data. We find values for the ratio $\rdsp^l \equiv {\mathrm{Br}(B\rightarrow \tau\nu_\tau D \pi)}/\mathrm{Br}(B\rightarrow l\nu_l D\pi)$ of $\rdsp^e=0.274\pm0.003$ and $\rdsp^\mu=0.275\pm0.003$, where the uncertainty comes from the uncertainty of the form factors parameters. Comparing against $R_{D\pi}$ reduces the gap with the latest LHCb result from $1.1\sigma$ to $0.48\sigma$, while the gap with the latest Belle result is reduced from $0.42\sigma$ to just $0.10\sigma$ and with the world average results from $3.7\sigma$ to $2.1\sigma$. Erratum added at the end of the file.
\end{abstract}

\maketitle

\pagestyle{myheadings}

\thispagestyle{empty}

\section{Introduction}

The formulation of the Standard Model (SM) incorporates the three families of leptons with universal couplings to gauge bosons, such that the differences in similar processes for different leptons are only of kinematical origin. This lepton flavor universality has been tested in the weak sector, for example, by studying the semileptonic decays of heavy mesons. The parameter that is expected to reflect the universality property is defined as the ratio of similar processes into two different leptons, namely
 \begin{equation}
 R_{X}\equiv\mathrm{Br}(P \to \tau\nu_\tau X)/ \mathrm{Br}(P \to l\nu_l X),
 \end{equation}
 where $P$ is the decaying particle, typically a pseudo-scalar meson, $ l=e, \mu$ and $X$ is the hadronic product. 

Early measurements of  $R_D$ \cite{BaBarOld1,BaBarOld2} from $B$ meson decays prompted the question of possible lepton flavour violation after finding significant discrepancies with the SM predictions, although a more recent measurement has obtained $R_D=0.375\pm0.064\pm0.026$ \cite{RDexp} which is in better agreement with the SM prediction of $R_D^{SM}=0.300\pm0.008$\cite{RDteo}.

\begin{figure}
\includegraphics[width=5cm]{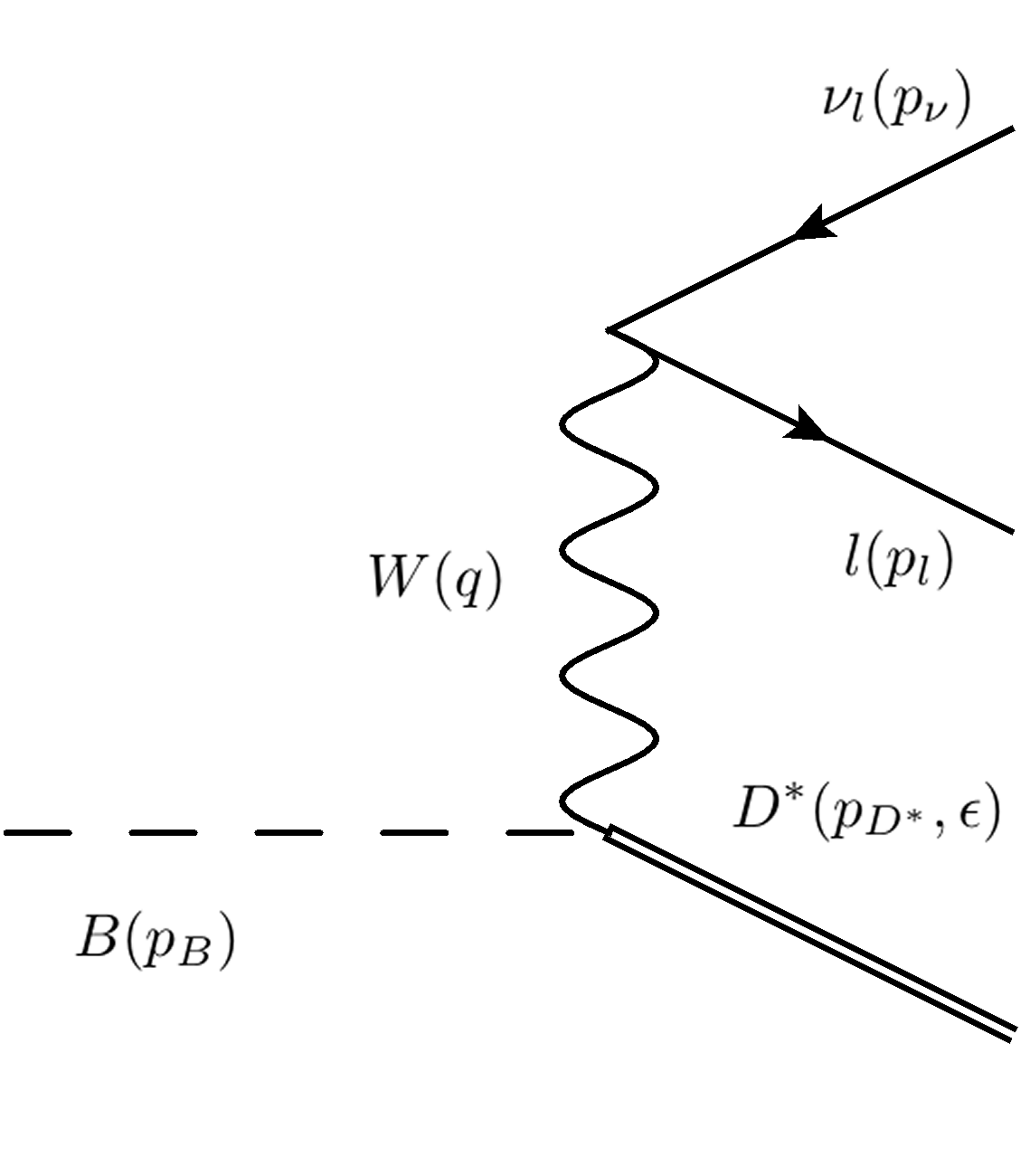}
\caption{Feynman diagram for $B\rightarrow l\nu D^*$ decay.}
\label{3b}
\end{figure}

For $R_{D^*}$, several B-factories conducted experiments which have also consistently measured values of $\rds$ that are higher than the SM prediction \cite{BaBarOld1,BelleOld,RDexp,LHCbOld,BaBarOld2}. The most recent results from the LHCb \cite{LHCbNewPrl,LHCbNewPrd} and Belle \cite{BelleNewPrd,BelleNewPrl} collaborations, that study the process through hadronic channels of the $\tau$ decay, have reduced this disagreement gap to a statistical significance of just $1.1 \sigma$ and $0.42\sigma$, respectively. However, a discrepancy with combined statistical significance of $3.7 \sigma$~\cite{Averages} is still found when considering the previous experiments which use other $\tau$ reconstruction methods \cite{BaBarOld1, BaBarOld2,RDexp,LHCbOld,BelleOld}. Since the deviation is still large in the $R_{D^*}$ case, we can inquire at which extent the nature of the hadronic particle in the final state affects the result. The $B\rightarrow l\nu D^*$  decay is represented by a single tree-level diagram, shown in Fig.~\ref{3b}. Since the $D^*$ decay width is relatively small, its decay process is usually considered not to play any role in the estimation of $\rds$ (since it is common to both leptonic decay modes, it should cancel in the ratio). Thus, it is common to consider simply this 3-body decay, from which a value of $\rds^{SM}=0.252\pm0.003$ is obtained \cite{VerticeBDW}. The error bar accounts for the uncertainties on the form factors of the vertex connecting the $B$, $W$ and $D^*$ particles as measured by Belle \cite{FormFactors}.\\
We argue that the tension of the theoretical prediction with experimental measurements is due in part to the fact that $\rds$ is not the proper quantity that the results should be compared to, since in all the experiments the $D^*$ is never measured directly but through its decay into daughter particles, namely a $D\pi$ pair for the charged $D^*$  (with a branching ratio of $98.4\%$)  and either a $D\pi$ pair (branching ratio of $64.7\%$) or $D\gamma$  (branching ratio of $35.3\%$) for the neutral $D^*$.
LHCb  and early Belle results on $\rds$ rely on purely charged $D^*$ \cite{LHCbNewPrl,LHCbNewPrd,BelleOld} (neutral B's)
while other Belle results \cite{RDexp,BelleNewPrl,BelleNewPrd} and Babar \cite{BaBarOld1,BaBarOld2} consider both neutral and charged ones. Therefore, it is adequate to consider the ratio
\begin{equation}
R_{D\pi} \equiv \mathrm{Br}(B\rightarrow \tau\nu_\tau D \pi)/\mathrm{Br}(B\rightarrow l\nu_l D\pi),
\end{equation}
obtained from the full 4-body diagram shown in Fig.~\ref{4b} as a better value to compare to.
Upcoming experiments will help to settle down the experimental value, and a solid SM prediction to compare to is mandatory.

In this work, we show that the corrections that arise from the full process, corresponding to adding the longitudinal degree of freedom of the off-shell $D^*$, are not negligible and help to get the experimental and theoretical results in better agreement. The corrections that we focus on in this work apply only to the $D\pi$ channel and the $D\gamma$ channel is ignored throughout.
An earlier work has considered the full process to explore the possible effect of other resonances and found it to be negligible \cite{kim}. The purely longitudinal contribution ratio has also been computed \cite{VerticeBDW}. Here we exhibit the role of each contribution, transverse, longitudinal and interference. In particular, the interference turns out to be very different for light and heavy leptons, from which $\rdsp$ is found to deviate from $\rds$.
 
In section \ref{2} we elaborate on the origin of these corrections and contrast with the shortened 3-body decay. Then, we present our result for $R_{D\pi}$ including the uncertainties estimate. The discussion and concluding remarks are presented in section \ref{3}.

\section{Longitudinal corrections. From 3 to 4 body decay}\label{2}
\begin{figure}
\includegraphics[width=6cm]{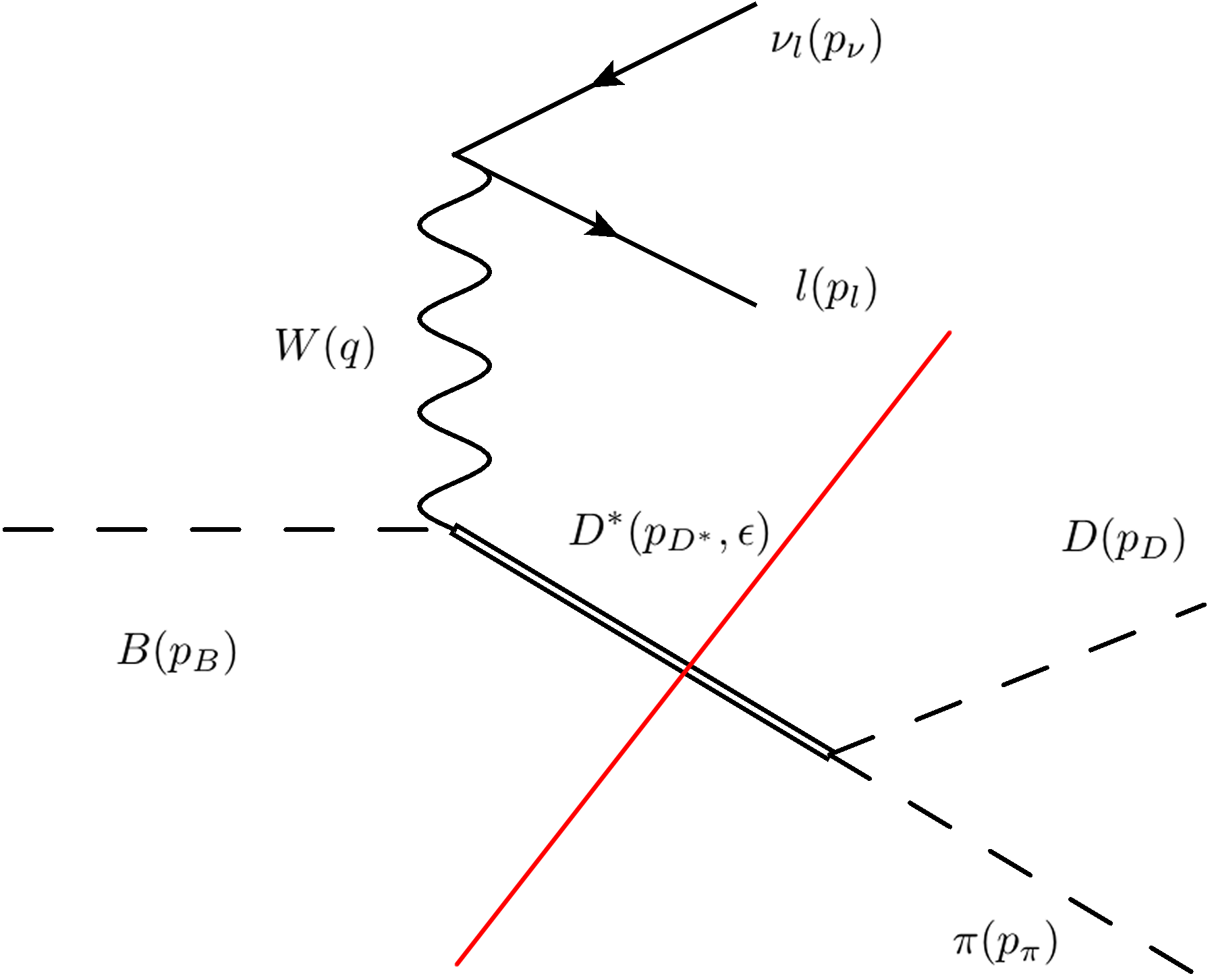}
\caption{Feynman diagram for the full $B\rightarrow l\nu_l D\pi$ decay, often thought of as independent $B\rightarrow l\nu D^*$ and $D^*\rightarrow D\pi$ decays as indicated by the red cutting line. }
\label{4b}
\end{figure}

Let us describe the 4-body decay process, taking  the 3-body one as a baseline to identify the role of the longitudinal contribution of the $D^*$ vector meson.\\
The $B\rightarrow l\nu D^*\rightarrow l\nu D\pi$ decay can be considered as separate subsequent processes by "cutting" the diagram as shown in Fig. \ref{4b}. This procedure kills the longitudinal contribution by considering the $D^*$  as on-shell.
In this approach, the total amplitude can be written as a product of 3-body and 2-body decay amplitudes connected by the outgoing and ingoing $D^*$ polarizations, and a sum over them to account for all the different forms they can match:
$$M=M_{3\mu}\left(\sum\epsilon^{*\mu}\epsilon^\nu\right) M_{2\nu}=M_{3\mu}\left(-g^{\mu\nu}+\frac{\pds^\mu\pds^\nu}{\dsm^2}\right) M_{2\nu},$$
where $\epsilon^\nu$, $\pds$ and $\dsm$ are the polarization tensor, momentum and mass of the $D^*$ respectively. $M_{3\mu}$ and $M_{2\nu}$ are the 3-body $D^*$ production and 2-body $D^*$ decay amplitudes, respectively, with the polarisation tensor factored out. The squared amplitude is then:
\begin{equation}\begin{split}
\label{3bamp}
|M|^2=M_{3\mu}M_{3\alpha}^*&\left(-g^{\mu\nu}+\frac{\pds^\mu\pds^\nu}{\dsm^2}\right)\\
\times &\left(-g^{\alpha\beta}+\frac{\pds^\alpha\pds^\beta}{\dsm^2}\right)M_{2\nu}M_{2\beta}^*.
\end{split}\end{equation}

The complete 4-body amplitude, on the other hand, is given by
$$M=M_{3\mu}D^{\mu\nu}M_{2\nu},$$
where $D^{\mu\nu}$ is the $D^*$ propagator which, upon considering the absorptive correction (dominated by the $D\pi$ mode as discussed before), can be set in terms of the transverse and longitudinal part as follows \cite{Wabsorptive,FiniteWidth, DsPropagator}:
\begin{equation}
\label{prop}
D^{\mu\nu}=\frac{-i T^{\mu\nu}}{\pds^2-\dsm^2+i Im\Pi_T}+\frac{i L^{\mu\nu}}{\dsm^2- i Im\Pi_L},
\end{equation}
with the corresponding projectors:
$$T^{\mu\nu}\equiv g^{\mu\nu}-\frac{\pds^\mu\pds^\nu}{\pds^2}~~~~~\mathrm{and}~~~~~L^{\mu\nu}\equiv \frac{\pds^\mu\pds^\nu}{\pds^2}.$$
 Here, the transversal correction is proportional to the full decay width, $Im\Pi_T=\sqrt{\pds^2} \Gamma_{D^*}(\pds^2)$, while the longitudinal function $Im \Pi_L$ is proportional to the $D-\pi$ mass difference, as will be discussed below.

Since $\Gamma_{D^*}\equiv\Gamma_{D^*}(\dsm^2)$ is relatively small, the relevant contribution to the transversal term is just around $\pds^2=\dsm^2$, and a narrow width approximation can be used. This allows us to rewrite the transversal part of the squared amplitude as
\begin{equation}\label{Tamp}
|M_T|^2=M_{3\mu}M_{3\alpha}^*T^{\mu\nu}T^{\alpha\beta}M_{2\nu}M_{2\beta}^*\frac{\pi \delta(\pds^2-\dsm^2)}{\dsm\Gamma_{D^*}}.
\end{equation}
We note that the delta function forces the transversal part of the $D^*$ to remain on-shell, thus recovering the same structure from (\ref{3bamp}), with global factors in (\ref{Tamp}) compensated by the splitting of the phase space in (\ref{3bamp}).
Therefore, taking only the transversal part of the $D^*$ propagator is equivalent to working with the on-shell scheme that is often used.

On the other hand, the longitudinal part of the propagator is not restricted to be around $\pds^2=\dsm^2$, and gives place to two new terms in the squared amplitude (one purely longitudinal and one of interference) that cannot be accounted for in the $B\rightarrow l\nu_lD^*$ process. 

The 2-body decay amplitude is $M_{2\nu}=-ig(p_D-p_\pi)_\nu$, where $p_D$ and $p_\pi$ are the momenta of the $D$ and the $\pi$, respectively, and g is the $D^*-D-\pi$ coupling. Thus, the longitudinal part of the amplitude can be written as
\begin{equation}
M_L=igM_{3\mu}\frac{\pds^\mu}{\pds^2}\frac{m_D^2-m_\pi^2}{\dsm^2-iIm\Pi_L}.
\end{equation}
Hence, the longitudinal corrections are modulated by a dimensionless mass-difference parameter $\Delta^2\equiv(m_D^2-m_\pi^2)/\dsm^2$. This is a known result for any vector meson that decays into two pseudo-scalars, and $\Delta^2$ is usually invoked as a suppression factor. However, due to the large mass difference between the $D$ and  $\pi$ mesons, here it happens to be a relatively large value of $\Delta^2=0.86$. Thus, the longitudinal corrections that are missing in the $B\rightarrow l\nu_lD^*$ case may carry an important weight.

The 3-body decay amplitude can be written as
\begin{equation}
M_3^\mu=\frac{G_F}{\sqrt 2}  V_{cb}J^{\lambda\mu} l_\lambda,
\end{equation}
where $l_\lambda\equiv\bar u_l\gamma_\lambda(1-\gamma^5)v_\nu$ is the leptonic current, $V_{cb}$ is the CKM matrix element and $J^{\lambda\mu}$ is the hadronic matrix element with the polarisation vector of the $D^*$ factored out. The hadronic matrix element can be parameterised in terms of four form factors \cite{NeubertR3b,neubertSO} (see appendix). The parameters of such form factors have been obtained from a heavy quark analysis of $B^0$ decays with electron and muon products measured by the Belle collaboration~\cite{FormFactors}.

In calculating the new decay widths, both the transversal and interference parts of the squared amplitudes have been integrated using the narrow width approximation (as the transversal part forces the main contribution from the interference to come from its corresponding dominant part), while the longitudinal part can be integrated without any restriction in the full 4-body phase space. In order to compare with the experimental determination, we restricted $D$ and $\pi$ momenta to fulfil  $(p_D+p_\pi)^2= (\dsm \pm \delta)^2$, where $\delta$ is in the range $\Gamma_{D^*}/2 $ to 1 MeV. Our results for $R_{D\pi}$ are the same in this range, as the small pure longitudinal correction is well below the current precision. The kinematics is taken as given in \cite{kumar} and implemented with the Vegas subroutine. We have found that the interference term makes a slight distinction between the $l=e$ and $l=\mu$ cases. Thus, we quote our final result separately as

$$R_{D\pi}^e=0.274\pm0.003$$
and
$$R_{D\pi}^\mu=0.275\pm0.003,$$

\noindent where the uncertainty comes from the uncertainties on the measurement of the form factors ($V$, $A_0$, $A_1$ and $A_2$) that characterize the hadronic vertex between the $B$, $D^*$ and $W$, for which we have used results published by Belle \cite{FormFactors} in agreement with world-average measurements \cite{Averages}. Notice that $A_0$ is not independent, but derived from $A_2$ as discussed in the appendix. Thus, an anti-correlation on the form factors is present, which brings the uncertainties lower.

In Table \ref{table1}, we show the contribution to the branching ratio from the transversal, longitudinal ($\delta=\Gamma_{D^*}$) and interference parts of the amplitude for all three lepton flavor products, which are consistent with Belle measurements for the electron and muon \cite{FormFactors}. Within parenthesis we quote the errors coming from the uncertainties in the form factor parameters and $V_{cb}$ \cite{FormFactors}.
We note that the pure longitudinal contribution to the branching ratio is the same for the light leptons. An early estimation of the contribution of the $D^*$ longitudinal polarization to the $B\rightarrow l\nu D^*$ process \cite{VerticeBDW} quotes a value for the pure longitudinal rate of $0.115(2)$, which is in agreement with our result of $0.111(3)$ after integration in the full 4-body phase space.
In the last two rows we show $R_{D\pi}$ for the electron and the muon as each part is added, namely, transversal, longitudinal and interference parts. Notice that, due to the cancellation of global factors in the ratio, $R_{D\pi}$ has a much higher precision than the individual branching ratios.

On the other hand, the relative size of the interferences turns out to increase with the lepton mass, being negligible for the electron and muon at the current precision, while for the tau it becomes of the order of 10\% compared to the transversal contribution. An earlier estimate \cite{kim} got a negligible effect from the interference that can be traced back to the propagator they used, which differs from ours (in the limit of $Im\Pi_L=0$) by a term proportional to $i\dsm \Gamma_{D^*}$ in the longitudinal part. The interference, upon integration in the narrow width approximation, leaves this term as the only not null contribution. Notice that this imaginary term makes a real contribution as both the leptonic tensor ($l^{\lambda \theta}=\sum l^\lambda l ^{\dagger\theta}$) and the $B-D^*-W$ vertex (see appendix) carry also an imaginary term.
This is the case, for example, when considering the contraction with the Levi-Civita from the leptonic tensor $\sum_{pol}l^\mu l^{\nu\dagger} $ and the term proportional to $A_1$ in the hadronic matrix element. We may interpret the interference as connecting the leptonic and hadronic parts mostly through the chirality of the lepton, which knows about its mass. Thus, the corrections in the hadronic part are modulated by the corresponding  leptonic flavour, making in this case the heavier the lepton the larger the contribution.

\begin{table}\caption{Contribution to the branching ratio of the transversal, longitudinal  ($\delta=\Gamma_{D^*}$) and interference parts of the amplitude for all three lepton flavors. Quantities are given in percentage. The last two rows shows the value of $R_{D\pi}$ as each contribution is added subsequently from left to right for $e$ and $\mu$.}\label{table1}\begin{tabular}{ c|c c c }
& Transversal & Longitudinal & Interference\\
\hline
Electron & $4.6(3)$ & $5.0(3)\times10^{-6}$ & $7.6(6)\times 10^{-8}$\\
Muon & $4.6(3)$ & $5.0(3)\times 10^{-6}$ & $1.6(1)\times10^{-3}$ \\
Tau & $1.16(8)$ & $1.1(6)\times10^{-6}$ &$ 1.02(7)\times10^{-1}$ \\
\hline
$R_{D\pi}^e$ &0.252&0.252&0.274\\
$R_{D\pi}^\mu$ &0.252&0.252&0.275
\end{tabular}\end{table}

\section{Discussion}\label{3}
We have argued that the experimental information for $\rds$ must be compared with the theoretical
$\rdsp$, since the experimental information relies on the reconstruction of the full 4-body decay process. We have shown that the longitudinal correction from the $D^*$ propagator introduces a correction to the branching ratios, which produces a value of $\rdsp^e=0.274\pm0.003$ and $\rdsp^\mu=0.275\pm0.003$, where the uncertainty comes from the experimental measurement of the form factor parameters. Within the error bars, $\rdsp$ can be considered as flavor-independent. This contrasts with the value of $\rds=0.252\pm0.003$ \cite{VerticeBDW} often used to compare with. The two quantities are distinguishable from each other at the current precision and therefore, the correction introduced is meaningful when comparing with the experimental result. Namely, by comparing with $\rdsp$ instead, we find that the difference with the latest results from LHCb \cite{LHCbNewPrl,LHCbNewPrd} goes down from $1.1\sigma$ to $0.48\sigma$, while the difference with the latest Belle results \cite{BelleNewPrl,BelleNewPrd} goes down from $0.42\sigma$ to just $0.10\sigma$, and the difference with the world average results \cite{Averages} goes down from $3.7\sigma$ to $2.1\sigma$. In all cases the agreement with the experiments is improved, but there still remains some tension with the world average results that cannot be explained by the longitudinal corrections alone.

\begin{figure}
\includegraphics[width=7cm]{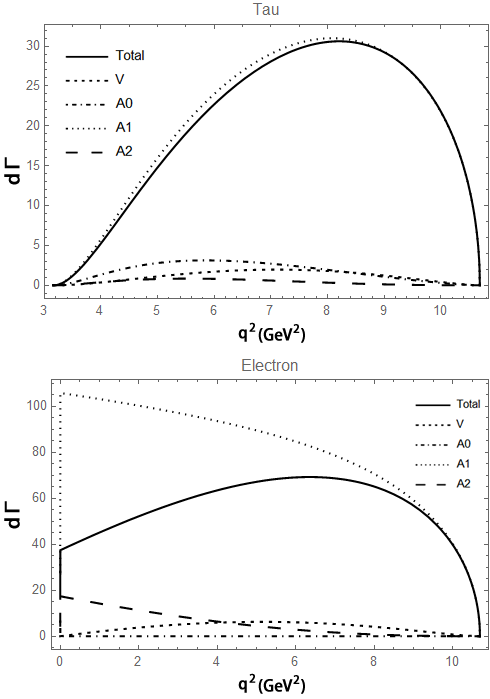}
\caption{Contribution to the three-body differential decay width of each form factor as a function of $q^2$, for the tau and the electron. For the electron, the $A_0$ term is nearly zero. Interference terms between form factors are not shown. }
\label{ff}
\end{figure}

In order to exhibit the role of the form factors, in Fig.~\ref{ff} we show the contribution of each of them to the three-body differential decay width for the case of the tau (upper panel) and electron (lower panel). Interferences are not shown.
The vector-like contribution $A_1$ is the dominant in each case, while the other contributions compete among themselves. Lattice calculations have provided information on this form factor at zero recoil which is consistent with the current experimental information \cite{latticeA1}.
The $A_0$ form factor accounts for the longitudinal projector for the transferred momentum $q$,  corresponding to states with helicity zero for the lepton-neutrino system. Because such a state is forbidden in the zero-mass limit for both the lepton and the neutrino, the term is heavily suppressed for both the electron and the muon, but not the tau. Measurements of the vertex have only been done for electric and muonic flavors~\cite{FormFactors, Averages}, where it is undetectable. Instead,
 $A_0$ is obtained from $A_2$ by invoking an approximate relation derived from heavy quark effective theory~\cite{VerticeBDW,NeubertR3,NeubertR3b} and used as such in the tau system. The drastically different behavior that $A_0$ has between flavors might be  important for understanding the experimental results. Thus, it is important to measure the form factors and in particular $A_0$ using tau decay channels, to shed light on the nature of possible deviations from the SM. 

Additional scalar resonances like the $D^*_0(2400)^0$ and longitudinal effects of the vector resonance together can be seen as background contributions. 
They might be removed from experimental data by subtracting a MC generated sample of the full scalar contribution \cite{DsPropagator}. In that case, the comparison of the experimental result is returned to be done with the 3-body transversal part, originally considered.
Additional tests of the lepton universality which are independent of the form factors will be useful to decide whether there is or not a deviation from the standard model \cite{lamn}. 

It has also been suggested that other observables defined in terms of the longitudinal contribution may be useful to search for new physics signals \cite{tanaka}. Early calculations on the second order corrections on the form factors \cite{neubertSO} and radiative corrections in the pseudo-scalar and neutral vector mesons \cite{radcor,radcor2,radcor3} systems have shown that they are expected to be important at the few percent level in the branching ratios. Thus, upcoming improvements on the experimental side will require to account for them in the theoretical prediction.

\begin{acknowledgments}
This work was supported in part by a PIIF-IFUNAM project and CONACyT project 252167F.
\end{acknowledgments}

\appendix*
\section{Hadronic vertex}
\label{vertex}

For the 3-body decay, the hadronic vertex connecting the $B$, $D^*$ and $W$ is characterized by four form factors as follows \cite{VerticeBDW}:
\begin{equation}\begin{aligned}
&\bra{D^*(\pds,\epsilon_\mu)}J^{\lambda}\ket{B(p_B)}=\label{vertex}\\ 
&\frac{2iV(q^2)}{m_B+\dsm}\epsilon^{\lambda \mu \alpha \beta}\epsilon^*_\mu(p_B)_\alpha (p_{D^*})_\beta 
-2\dsm A_0(q^2)\frac{q^\lambda q\cdot\epsilon^*}{q^2}\\
&-(m_B+\dsm)A_1(q^2)\left( \epsilon^{* \lambda} - \frac{q^\lambda q\cdot\epsilon^*}{q^2}\right)\\
&+
\frac{A_2(q^2) q\cdot\epsilon^*}{m_B+\dsm}
\left( (p_B+\pds)^\lambda - \frac{m_B^2-\dsm^2}{q^2} q^\lambda\right),
\end{aligned}\nonumber
\end{equation}
where $J^\lambda $ is the weak current, $\epsilon_\mu^*$ the polarization vector of the $D^*$, and $q=p_B-\pds$ the transferred momentum. These terms constitute the most general double-indexed Lorentz structure that can be constructed with the available variables, for an on-shell $D^*$. 
For the 4-body decay, as an approximation, we have used the same hadronic vertex and replaced the polarisation vector by a free index to contract with the hadronic part associated to the $D^*$ decay. In this approximation, terms proportional to $\pds^\mu$ that can coupled to the longitudinal part of the $D^*$ propagator are missing. Since the form factors have been derived from Belle data \cite{FormFactors} without including these terms, a new analysis should be necessary. Additional longitudinal terms in the $D^*-D-\pi$ vertex have not been considered either. A full 4-body description would require to consider the most general structure for the four body decay, similar to $K_{l4}$ as discussed in \cite{kl4}  where the form factors were computed in the chiral approach. It would be useful to have  them  computed at the heavy quark limit and isolate the $D^*$ contribution from other possible resonances that may be part of the internal process.

\newpage
ERRATUM: After publication we found a bug in the transcription code for the numerical integration. Here we present the corrected values for Table 1. Thus, the results on $R_{D\pi}$ are consistent with $\rds$ at the current precision (uncertainties from the form factors are $\pm 0.003$). This modifies our previous finding that lead us to a different conclusion. \\

\begin{table}[b]
\caption{Update of Table 1. Same caption.}
\label{table1new}
\begin{tabular}{ |c|c c c |} 
 \hline
& Transversal  & Longitudinal & Interference\\
\hline
Electron & $4.6(3)$  & $5.0(3)\times10^{-6}$ & $3.2(2)\times 10^{-3}$\\
Muon & $4.6(3)$ & $5.0(3)\times 10^{-6}$ & $3.2(2)\times10^{-3}$ \\
Tau & $1.16(8)$ & $1.1(6)\times10^{-6}$ &$ 6.4(2)\times10^{-4}$ \\
\hline
\hline
$R_{D\pi}^e$ &0.25221&0.25221&0.25217\\
$R_{D\pi}^\mu$ &0.25330&0.25330&0.25327\\
\hline
\end{tabular}
\end{table}

\bibliography{lfu}

\end{document}